# Efficient excitation of dye molecules for single photon generation

Ross C. Schofield,[1] Kyle D. Major,[1] Samuele Grandi,[1] Sebastien Boissier,[1] E. A. Hinds,[1] and Alex S. Clark[1,*]

[1] *Centre for Cold Matter, Blackett Laboratory, Imperial College London, Prince Consort Road, SW7 2AZ London, UK*

A reliable photon source is required for many aspects of quantum technology. Organic molecules are attractive for this application because they can have high quantum yield and can be photostable, even at room temperature. To generate a photon with high probability, a laser must excite the molecule efficiently. We develop a simple model for that efficiency and discuss how to optimise it. We demonstrate the validity of our model through experiments on a single dibenzoterrylene (DBT) molecule in an anthracene crystal. We show that the excitation probability cannot exceed 75% at room temperature, but can increase to over 99% if the sample is cooled to liquid nitrogen temperature. The possibility of high photon generation efficiency with only modest cooling is a significant step towards a reliable photon source that is simple and practical.

## I. INTRODUCTION

Many applications have need of a fast, reliable source of individual photons on demand [1]. These range from quantum sensing [2] through quantum communication [3] to full-scale photonic quantum computing [4]. The standard way to obtain single photons is to make a pair by parametric down conversion or spontaneous four-wave mixing, and use the detection of one photon to herald the presence of the other. The pair production occurs at random times, meaning the source has to run at a low rate to ensure only one pair is created at a time, so the simultaneous production of, say, 10 pairs from 10 separate sources is not a practical possibility [5]. There has been some progress in multiplexing a large number of heralded photon sources with suitable delays, but this approach still presents technological challenges [6–9]. A promising solution to this problem is to excite a single quantum system with a trigger pulse, then collect the photon that is spontaneously emitted. This is sometimes called a source of photons on demand. An isolated atom provides a particularly well defined quantum system [10, 11], but the need for ultra-high vacuum and some sort of trap makes this difficult to scale up in practical devices. Consequently, there is great interest in solid-state alternatives [12] such as quantum dots [13–16], defects in diamond [17, 18], impurities in other solids [19], and our system of choice – single organic molecules [20, 21]. If the photons are required to interfere with each other, they should be identical and then the source needs to be cooled to very low temperature to suppress spectral broadening due to thermal phonons. However, there are a number of applications where broadband single photons can still be useful, such as quantum imaging [22], quantum communication [3], or other applications requiring a number-squeezed light source.

In this paper we consider the use of a single dye molecule as a source of individual photons, produced by spontaneous emission after laser light excites the molecule. In particular, we investigate the use of polycyclic aromatic hydrocarbons (PAH) because these can have a radiative yield very close to 1 [23, 24]. Using a simple rate model to describe the steady, excited-state population induced by the laser, we discuss how to maximise that population so that the molecule is most likely to yield a photon. We demonstrate the validity of our model by experiments on a single dibenzoterrylene (DBT) molecule in an anthracene crystal, and we show that high photon generation efficiency can be achieved when the sample is cooled to liquid nitrogen temperature. This is a significant step in making a fast, reliable photon source that is simple and practical.

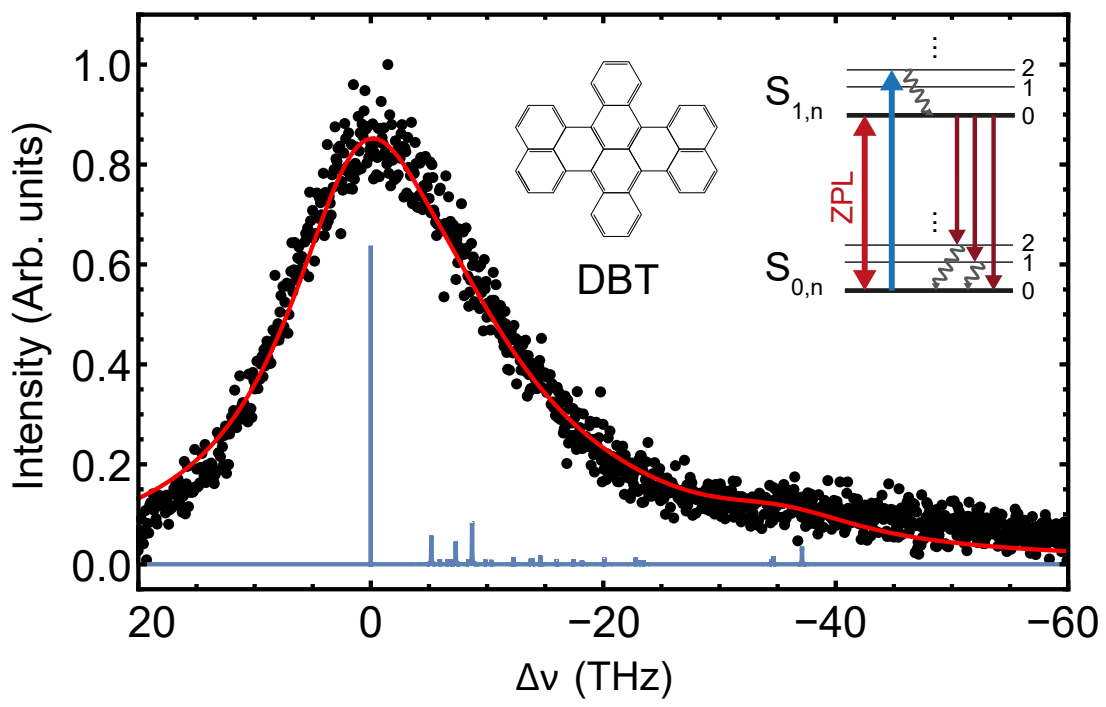

Figure 1. Dispersed emission spectrum of a DBT molecule. Black dots: spectrum measured at room temperature. Blue line: Analytical spectrum (see Eq.(8)) constructed to reproduce the low-temperature emission spectrum in Fig. 1(c) of Trebbia *et al.* [25]. Red line: the same analytical spectrum, but with $\Gamma 2_n$ increased to fit the our room temperature data. Inset: Molecular structure of DBT and generic energy level diagram showing pumping on the ZPL (red double-ended arrow), to a higher vibrational level (blue up arrow), fast non-radiative decays (black curvy arrows) and decays to higher vibrational levels of the ground state (maroon down arrows).

* alex.clark@imperial.ac.uk

## II. MODELLING THE MOLECULAR EXCITATION

The diagram inset in Fig. 1 illustrates the energy levels of a suitable molecule. The ground electronic state is a singlet, $S_0$, whose vibrational sublevels are designated $S_{0,n}$, $n$ being a label for the vibrational excitations. The first excited singlet, $S_1$, has sublevels $S_{1,n}$. The optical excitation $S_{0,0} \rightarrow S_{1,0}$ is followed by stimulated decay back to $S_{0,0}$, and by spontaneous decay in a few ns to the $S_{0,n}$ levels. Any population in $S_{0,n>0}$ relaxes in a few ps to $S_{0,0}$. Alternatively, the molecule may be excited from $S_{0,0}$ to $S_{1,n>0}$. From here the molecule relaxes (again in ps) to the $S_{1,0}$ state, from which it decays spontaneously to the $S_{0,n}$ levels. It is also possible for $S_1$ to decay to the lowest-lying triplet state, but for efficient fluorophores this decay branch is weak [26, 27] and therefore we neglect it in the present discussion.

We can reasonably require that the rate for excitation be much slower than the (GHz) rate for relaxation of the vibrational excitations. Then an excitation to $S_{1,n>0}$ is followed promptly by an incoherent transfer of population to the state $S_{1,0}$, whilst the subsequent decay to $S_{0,n}$ is immediately followed by relaxation to $S_{0,0}$. Writing the populations of $S_{1,0}$ and $S_{0,0}$ as $p_e$ and $p_g$ respectively, it is then a good approximation to consider that $p_g + p_e = 1$ and the excited state population obeys the rate equation

$$\frac{dp_e}{dt} = p_g \sum_{n=0}^{n_{max}} R_n - p_e(R_0 + \Gamma 1_0)\,. \tag{1}$$

Here $R_n$ is the stimulated transition rate on the transition $S_{0,0} \leftrightarrow S_{1,n}$, and $\Gamma 1_0$ is the spontaneous decay rate of the population $p_e$. The sum goes up to the highest vibrational level labeled by $n_{max}$.

After sufficient time, the populations relax to a steady state, where $\frac{dp_e}{dt} = 0$ and therefore

$$p_e = \frac{\sum_n R_n}{R_0 + \sum_n R_n + \Gamma 1_0}\,. \tag{2}$$

With strong excitation, such that $R_0 + \sum_n R_n \gg \Gamma 1_0$, $p_e$ saturates to the value

$$p_{e,\infty} = (1 + \frac{R_0}{\sum_n R_n})^{-1}\,. \tag{3}$$

If the only significant excitation is on the ZPL, $p_e$ saturates to $\frac{1}{2}$. If instead the excitation is entirely to states $S_{1,n>0}$, then $R_0 = 0$ and $p_e$ saturates to 1. This reproduces the well-known result that the steady state of a driven, damped 3-level system can have a population inversion, whereas the damped 2-level system cannot. From the perspective of building a triggered photon source, we want a trigger pulse to produce a steady state with the largest $p_e$ it can, and to do so quickly enough that the resulting spontaneous emission delivers a photon with high probability and well-defined timing. This argues for excitation with a sub-nanosecond pulse, and with a low rate $R_0$. It should be noted that pulsed excitation with a pulse length much longer than the coherence time in these systems ($\sim$ 20 fs) achieves the same $p_{e,\infty}$ as continuous-wave (cw) excitation, and these are the type we consider. As we now show, that requires excitation on the blue side of the ZPL, together with a low enough transverse relaxation rate. The excitation rates $R_n$ in Eq.(3) are found by solving the optical Bloch equations in the steady state:

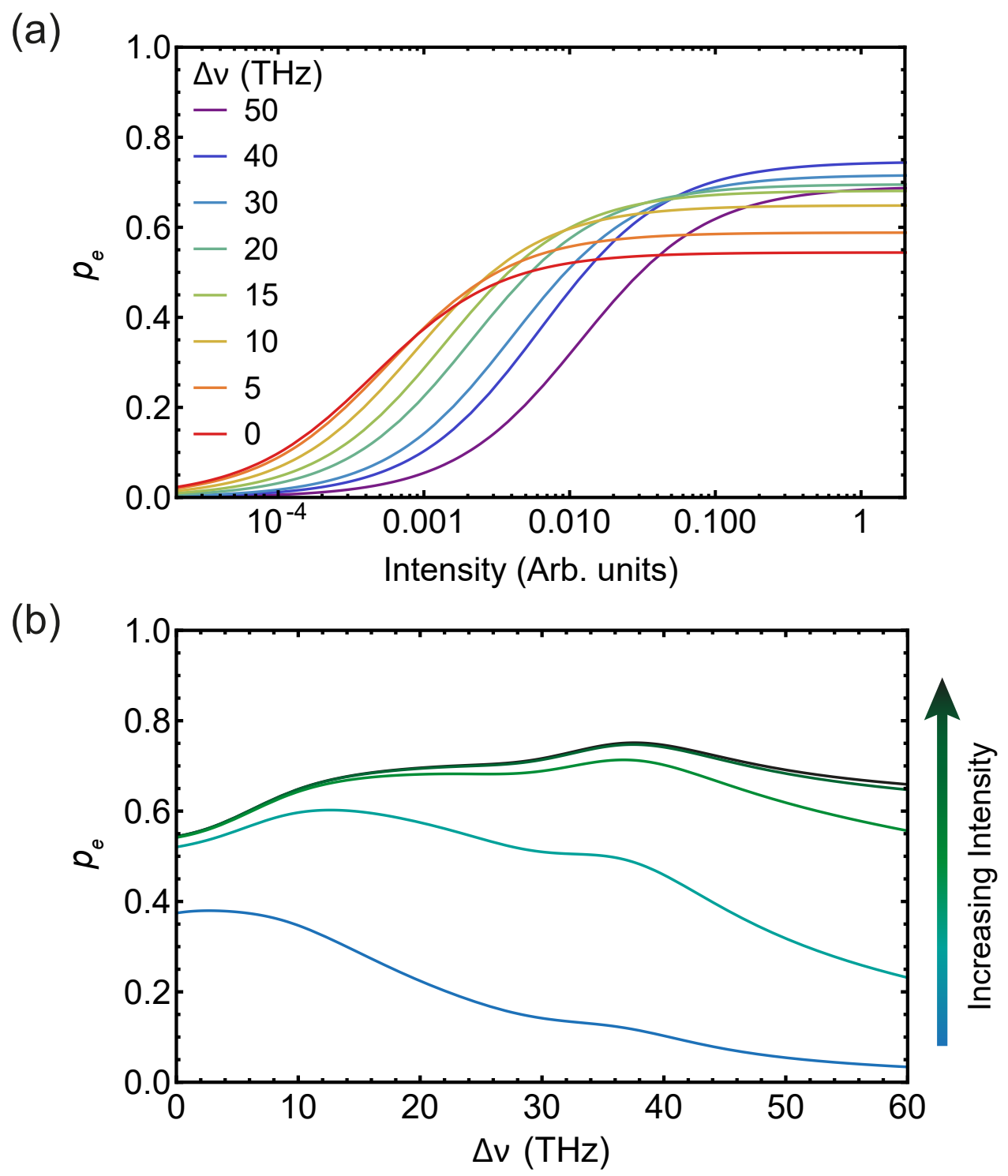

Figure 2. (a) Calculated excited state population for DBT in equilibrium at room temperature *versus* intensity of excitation laser for several detunings to the blue of the ZPL. (b) The same data plotted against detuning for several values of pump laser intensity, equivalent to the evolution of the absorption spectrum as the pump laser intensity is increased.

$$R_n = \frac{1}{2}\Omega_n^2 \frac{\Gamma 2_n}{\delta_n^2 + \Gamma 2_n^2}\,, \tag{4}$$

where $\Omega_n$, $\delta_n$ and $\Gamma 2_n$ are respectively the Rabi frequency, laser (angular) detuning from resonance, and transverse relaxation rate for the transition $S_{0,0} \leftrightarrow S_{1,n}$ [28, 29]. The Rabi frequency is

$$\Omega_n = \frac{1}{\hbar}|\vec{d}_n \cdot \vec{E}|\,, \tag{5}$$

where $\vec{d}_n = \langle S_{0,0}|\vec{d}|S_{1,n}\rangle$ is the dipole transition matrix element and $\vec{E}$ is the electric field driving the transition. Since all the transitions are driven by the same component, $E_{\parallel}$, of the light field, we have

$$R_n = \frac{I}{\epsilon\epsilon_0\hbar^2 c} d_n^2 \frac{\Gamma 2_n}{\delta^2 + \Gamma 2_n^2}\,, \tag{6}$$

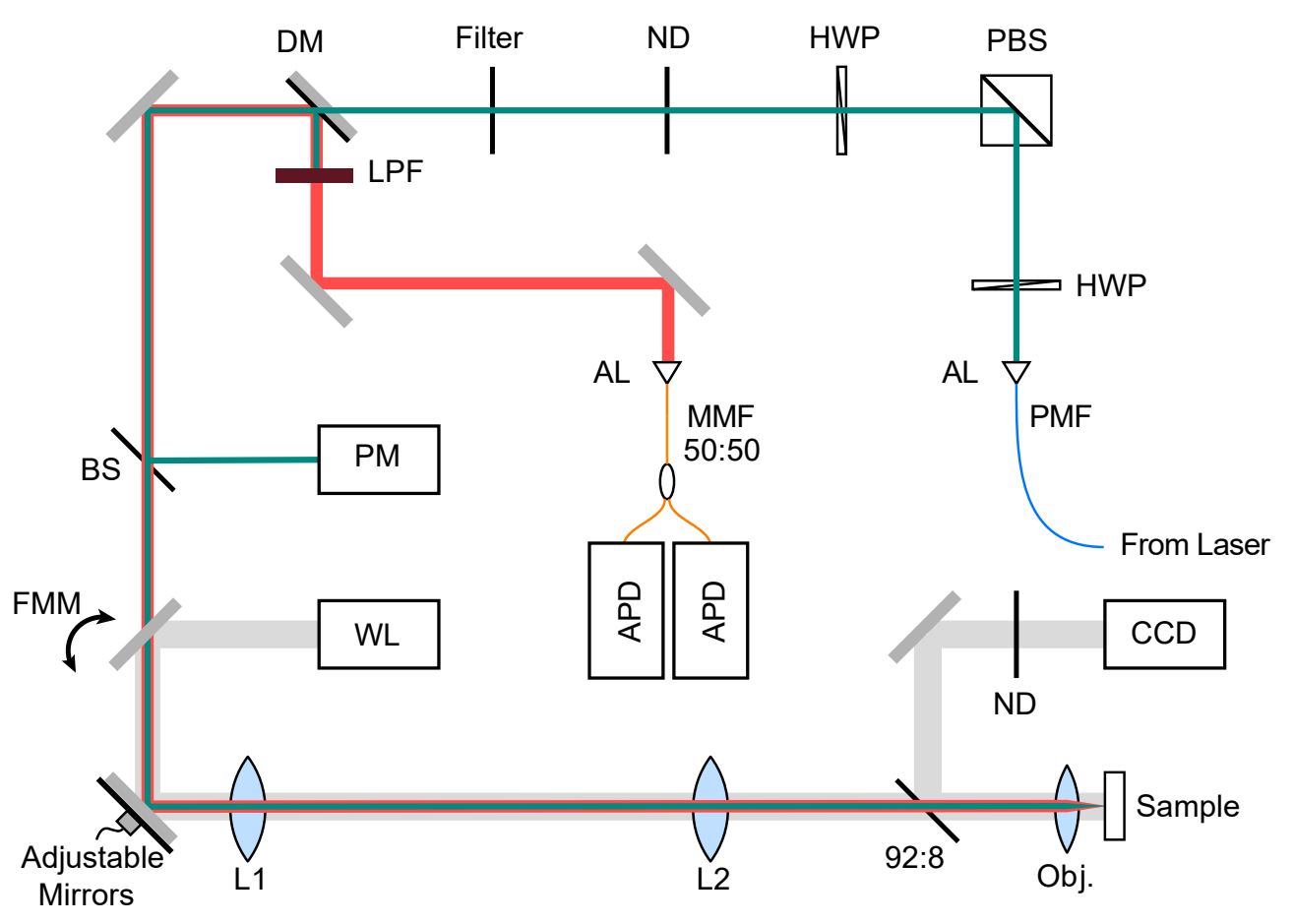

Figure 3. Schematic diagram of the confocal microscope. Dark green beam indicates the pump light, red is the fluorescence and grey is the white light used for imaging. PMF: polarization maintaining fibre; AL: aspheric lens fibre coupler; HWP: half-wave plate; PBS: polarizing beam splitter; ND: neutral density filter; DM: dichroic mirror; BS: beam sampler; PM: power meter; FMM: flip-mount mirror; WL: white light source; L1: first lens; L2: second lens; Obj: microscope objective lens; CCD: charge-coupled device camera; LPF: 800 nm long pass filter; MMF 50:50: multimode fibre beam splitter; APD: avalanche photodiode.

where the effective plane-wave intensity at the molecules, defined as $I = \frac{1}{2}\epsilon\epsilon_0 E_\parallel^2 c$, is proportional to the laser light intensity, $I_{ext}$, incident on the sample. Equation (6) allows us to evaluate Eq.(3) once we know the dipole matrix elements, the detunings and the transverse relaxation rates. In the following sections, we test this model in the specific case of a dibenzoterrylene (DBT) molecule in anthracene, but it could be equally well applied to other organic molecules with known dephasing rates and vibrational levels (see Supplementary Information).

### A. A specific case: dibenzoterrylene in anthracene

Many PAH molecules are fluorescent in the visible and DBT, with the formula $C_{38}H_{20}$ and the structure shown inset in Fig. 1, is no exception. Pure DBT appears green as it absorbs both blue and red light, but it is the red absorption from 700-800 nm, that is of interest here. Anthracene makes an excellent host because it is transparent at 700-800 nm, and all its electronically excited levels lie above above the $S_1$ levels of DBT, making it impossible for the DBT to exchange electronic excitation with its host [26, 29–32]. Moreover, DBT is highly photostable at room temperature when it is embedded in a crystal of anthracene [33, 34]. When DBT decays from the $S_{1,0}$ level, its fluorescence spectrum has Stokes sidebands due to the transitions $S_{1,0} \to S_{0,n}$, whose transition moments are $\tilde{d}_n = \langle S_{1,0}|d|S_{0,n}\rangle$. Supplementary Information lists the frequencies $\omega_n$ and relative intensities $\eta_n$ of these lines, taken from the low-temperature fluorescence spectrum in Fig. 1(c) of Trebbia *et al.* [25]. From these we deduce the relative strengths of the transition moments: $\tilde{d}_n^2 \propto \eta_n/\omega_n^3$. In these types of molecule, excitation of the optically active electron has very little influence on the vibrating bonds, and therefore the vibrational frequencies and wavefunctions are similar in the upper and lower electronic states [35]. We therefore make the reasonable approximation that the Franck-Condon factors are the same for the excitation $S_{0,0} \to S_{1,n}$ and the decay $S_{1,0} \to S_{0,n}$, and hence that $d_n$ is equal to $\tilde{d}_n$. In this way, we use the measured fluorescence spectrum to give us a set of (relative) values for the $d_n$ to be used in Eq.(6):

$$d_n^2 \simeq \tilde{d}_n^2 \propto \frac{\eta_n}{\omega_n^3}\,. \tag{7}$$

We turn now to the damping rates to be used in Eqs.(2) and (6). Previous measurements [29] give the population decay rate $\Gamma 1_0/(2\pi) \simeq 40\,\mathrm{MHz}$, corresponding to a spontaneous decay lifetime of 4 ns, and this remains the same whether at room temperature or at low temperature [36]. The damping rates $\Gamma 2_n$ in Eq.(6) describe relaxation of the optical coherences $|S_{1,n}\rangle\langle S_{0,0}|$. At 4 K, $\Gamma 2_n$ is small enough that one can excite a single $S_{1,n}$ level and observe fluorescence proportional to that $R_n$ alone, with a spectral width (FWHM) of $2\,\Gamma 2_n$. Using the strong $n = 7$ line 8.7 THz to the blue of the ZPL (see Fig. 1), we have measured a value $\Gamma 2_7/(2\pi) = 23(3)$ GHz (see Supplementary Information), corresponding to a decay time for the vibrationally excited population of $\sim 4$ ps. For simplicity we assume that all the $S_{1,n>0}$ states have similar vibrational relaxation times (the exact values are not critical for our purpose here).

At room temperature the optical coherence is so rapidly dephased by the interaction with thermal phonons that $\Gamma 2_n$ is a thousand time larger and the absorption lines are no longer resolved. Indeed, we are unable to measure the width of the room-temperature absorption spectrum directly because the excitation laser does not scan far enough. Instead, we measure the dispersed fluorescence spectrum while exciting continuously at 730 nm, with the result shown by the black dots in Fig. 1. We use this spectrum to determine the room-temperature value of $\Gamma 2_n$, as follows.

The blue curve in Fig. 1 shows the narrow lines of the dispersed fluorescence spectrum, $\mathcal{S}_{flu}$, at 2 K,

$$\mathcal{S}_{flu} = \sum_n \eta_n \frac{\tilde{\Gamma 2}_n^2}{(\omega-\omega_n)^2 + \tilde{\Gamma 2}_n^2}\,, \tag{8}$$

in which $\omega$ is the frequency of the emitted light and $\tilde{\Gamma 2}_n$ are the relaxation rates of the coherences $|S_{1,0}\rangle\langle S_{0,n}|$. We have taken $\tilde{\Gamma 2}_0 = \Gamma 1_0/2$ and $\tilde{\Gamma 2}_{n>0}/(2\pi) = 23\,\mathrm{GHz}$ on the assumption that all the vibrational excitations relax at roughly the same rate. The broad spectrum (red line) is obtained by adding a further rate $\tilde{\Gamma 2}^*$ to these low temperature rates in order to model the dephasing of the optical dipole by thermal phonons. On fitting

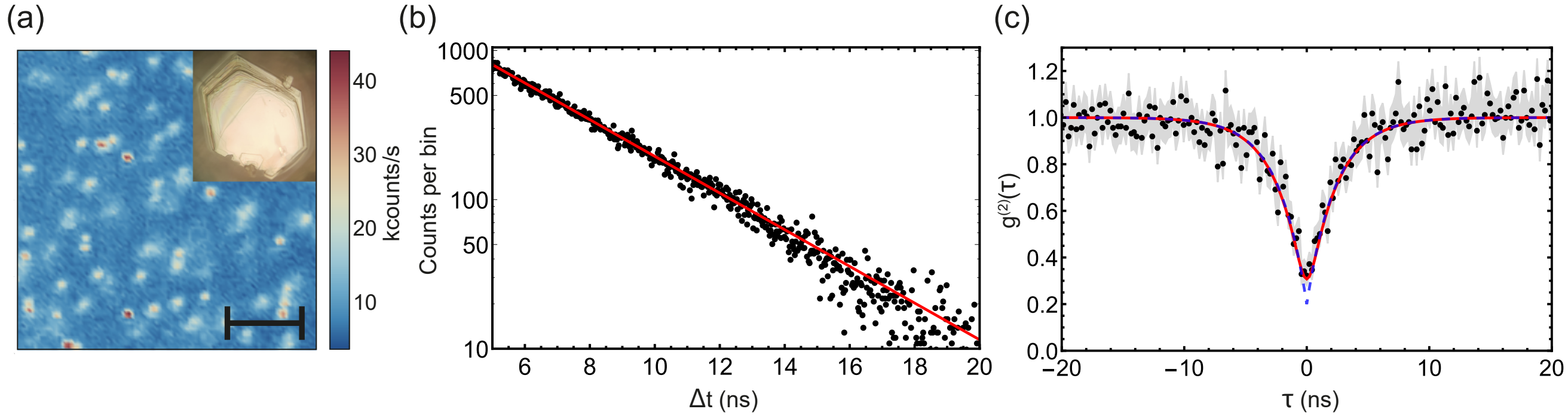

Figure 4. (a) Confocal microscope image showing fluorescence from individual DBT molecules in an anthracene crystal. The scale bar is 5 $\mu$m. Inset: white light image of the crystal, which is $\sim 0.5$ mm across. (b) Histogram of time delays, $\Delta$t, between pulsed excitation of a single molecule and subsequent detection of a fluorescence photon. The bin size is 26.7 ps. Red line: single exponential fit to the data, giving excited-state lifetime 3.53(2) ns. (c) Black dots: measured $g^{(2)}(\tau)$ for the same molecule, with grey shading indicating $1\sigma$ Poissonian uncertainty. Blue dashed line: exponential fit to the data. Red curve: fit using Eq.(9) convolved with a Gaussian to model timing jitter of the detectors

this curve to the data we obtain the value $\tilde{\Gamma 2}^*/(2\pi) = 8.2\,\mathrm{THz}$. We expect the coherence $|S_{1,n}\rangle\langle S_{0,0}|$ to suffer the same thermal dephasing and therefore we conclude that $\Gamma 2_n/(2\pi) = 8.2\,\mathrm{THz}$ at room temperature.

With these values of $d_n^2$, $\Gamma 2_n$ and $\Gamma 1_0$, we can substitute Eq.(6) into Eq.(2) to give the excited state population $p_e$ prepared by the trigger pulse. Figure 2(a) is a plot of $p_e$ at room temperature against the intensity of the pulse, for several blue detunings $\Delta\nu$ of the laser from the ZPL. As anticipated in Eq.(3), $p_e$ saturates at various levels $p_{e,\infty}$, the lowest being 0.55 when $\Delta\nu = 0$ and the highest being 0.75 when $\Delta\nu = 40\,\mathrm{THz}$. At 50 THz detuning $p_{e,\infty}$ decreases again because there are only 23 vibrationally excited states in the model, and indeed no more have been observed experimentally. Thus, it seems likely that $p_{e,\infty} = 0.75$ is the highest achievable at room temperature. Figure 2(b) shows the same information, but plotted against detuning for several intensities, indicated by the colour of the lines.

## III. DBT MOLECULE AT ROOM TEMPERATURE

We now describe experiments to test this theory on a single DBT molecule at room temperature.

### A. Experimental setup

Individual DBT molecules were observed using the confocal microscope shown schematically in Fig. 3. The molecule is excited either by a cw tuneable titanium sapphire laser (Coherent MBR) or by a pulsed diode laser centred at 781 nm (Picoquant). The laser light is coupled to the table through a polarisation maintaining fibre (PMF), and is collimated using an aspheric lens (AL). A half wave plate (HWP) and polarising beam splitter (PBS) clean the linear polarisation of the fibre output, then a second HWP rotates the linear polarisation to lie along the optical transition dipole of the DBT molecule. A neutral density filter (ND) controls the pump power and a filter (Filter) removes any light of unwanted wavelength coming from the laser or from fluorescence of optical components up to this point. This is a $780 \pm 6$ nm band pass, a $760 \pm 6$ nm band pass, or a 750 nm short pass, depending on the chosen laser wavelength. After passing through the dichroic mirror (DM) and beam sampler (BS) (used to monitor the forward power), the angle of the beam is scanned by two galvo mirrors. The beam then passes through lenses, L1 (focal length $f_1 = 75$ mm) and L2 ($f_2 = 250$ mm), in a '4f' arrangement, where the scanning mirrors lie at $2f_1 + 2f_2$ from the back focal plane of the microscope objective (Obj, a Nikkon PlanApo 100x, 0.9NA). This arrangement allows the galvo mirrors to scan the focal spot on the sample, while always filling the aperture of the objective to produce a Gaussian spot of $\sim 720$ nm FWHM. The DBT-doped anthracene crystal, shown inset in Fig. 4(a), is grown by co-sublimation [27], then placed on a glass coverslip and protected by a thin, spin-coated layer of polyvinyl alcohol (PVA). This sample is held in place using a vacuum chuck. Red-shifted fluorescence from the DBT molecules travels back to the dichroic mirror, which reflects it through an 800 nm long pass filter (LPF, to block scattered excitation light), and into a multimode fiber (MMF). A 50:50 fiber splitter delivers the light to two avalanche photodiodes (Perkin Elmer, APD) so that the second-order correlation function, $g^{(2)}(\tau)$, can be measured.

### B. Single molecule images

The microscope image in Fig. 4(a) shows fluorescence from isolated DBT molecules over a 20 $\mu$m square area of the anthracene crystal. We selected a single molecule, illuminated it with the 781 nm pulsed laser, and monitored the time $\Delta t$ between excitation and the detection of photons to build up the histogram shown in Fig. 4(b). Fitting the decay curve to a single exponential decay $Ae^{-\Gamma 1_0 \Delta t}$, shown as a red line, we measured the excited state lifetime $1/\Gamma 1_0 = 3.53(2)$ ns.

Changing to $\sim 70\,\mu$W of cw laser light at 780 nm, and using both APD detectors, we measured $g^{(2)}(\tau)$, plotted in Fig. 4(c). This shows a strong dip going down to 0.33(4) at $\tau = 0$. With the high dephasing rate at room temperature, the $S_{1,0}$ level is populated incoherently, and thefore the ideal second order correlation function takes the simple form [29]

$$g^{(2)}(\tau) = 1 - \frac{1}{N_{\text{eff}}} e^{-(1+S)\Gamma 1_0 |\tau|}\,, \qquad (9)$$

where $N_{\text{eff}}$ is the effective number of molecules contributing to the signal, $S = I/I_{\text{sat}}$ is the saturation parameter, and $\tau$ is the time delay between the first and second detection event. The blue dashed curve in Fig. 4(c) is a fit to this function with $\Gamma 1_0$ fixed at our measured value. That works well in the wings but has a discrepancy near $\tau = 0$ because of the 455 ps (rms) timing jitter of our detection system. When we convolve Eq.(9) with a Gaussian to represent the jitter, we obtain the red curve which gives $S = 0.5$ and $N_{\text{eff}} = 1.25$. The corresponding de-convolved $g^{(2)}(0)$ is 0.2. Further, there is a background – about 10% of the peak molecule count rate found from a region of anthracene with no molecule present in a confocal scan – that generates some accidental coincidences. After correcting for that [32], we conclude that $N_{\text{eff}} = 1.02(4)$ and hence that we are indeed looking at a single molecule.

### C. Scattering rate at room temperature

Having characterised this particular molecule, we investigated the saturation of its scattering rate by analysing the confocal microscope images. The intensity distribution of the laser spot was a slightly elliptical Gaussian with principal axes along the two axes of the scan and with standard deviations $\sigma_x$ and $\sigma_y$. In Eq.(2), the rates $R_n$ are all proportional to the laser intensity, so a molecule at position $(x_0, y_0)$, illuminated by a spot centred on $(x, y)$, gives a signal

$$R(x,y) = R_\infty \frac{S_0}{S_0 + \exp\left[\frac{1}{2}\left(\frac{x-x_0}{\sigma_x}\right)^2 + \frac{1}{2}\left(\frac{y-y_0}{\sigma_y}\right)^2\right]}\,, \qquad (10)$$

where $R_\infty$ is the fully saturated rate of photons detected from the molecule and $S_0 = I_0/I_{\text{sat}}$ is the saturation parameter when the spot is centred on the molecule. When $S_0 \ll 1$ this gives an image that reproduces the intensity distribution of the laser spot, but as the molecule saturates, the image becomes wider and flatter.

With the excitation laser frequency set at a blue detuning $\Delta\nu$ from the ZPL we took microscope images over a range of intensities and made a global fit of Eq.(10) to these images, to obtain a value for $R_\infty$ at that detuning (see also Supplementary Information). This was done for 8 different values of $\Delta\nu$ with the results plotted in Fig. 5(a). It is straightforward to convert these to values of $p_{e,\infty}$ through the relation $R_\infty = \alpha\, p_{e,\infty}\, \Gamma 1_0$, where $\alpha$ is the total efficiency for collection and detection. On fitting Eq.(3) to the data in Fig. 5(a), with $\alpha$ as the only free parameter, we find the result $\alpha = 1.03(2) \times 10^{-3}$ for this particular molecule.[1] The correspondence between our data and the theory gives us confidence that Eqs.(3) and (6) provide a correct description, and therefore that one cannot expect to achieve more than 75% steady-state excitation at room temperature.

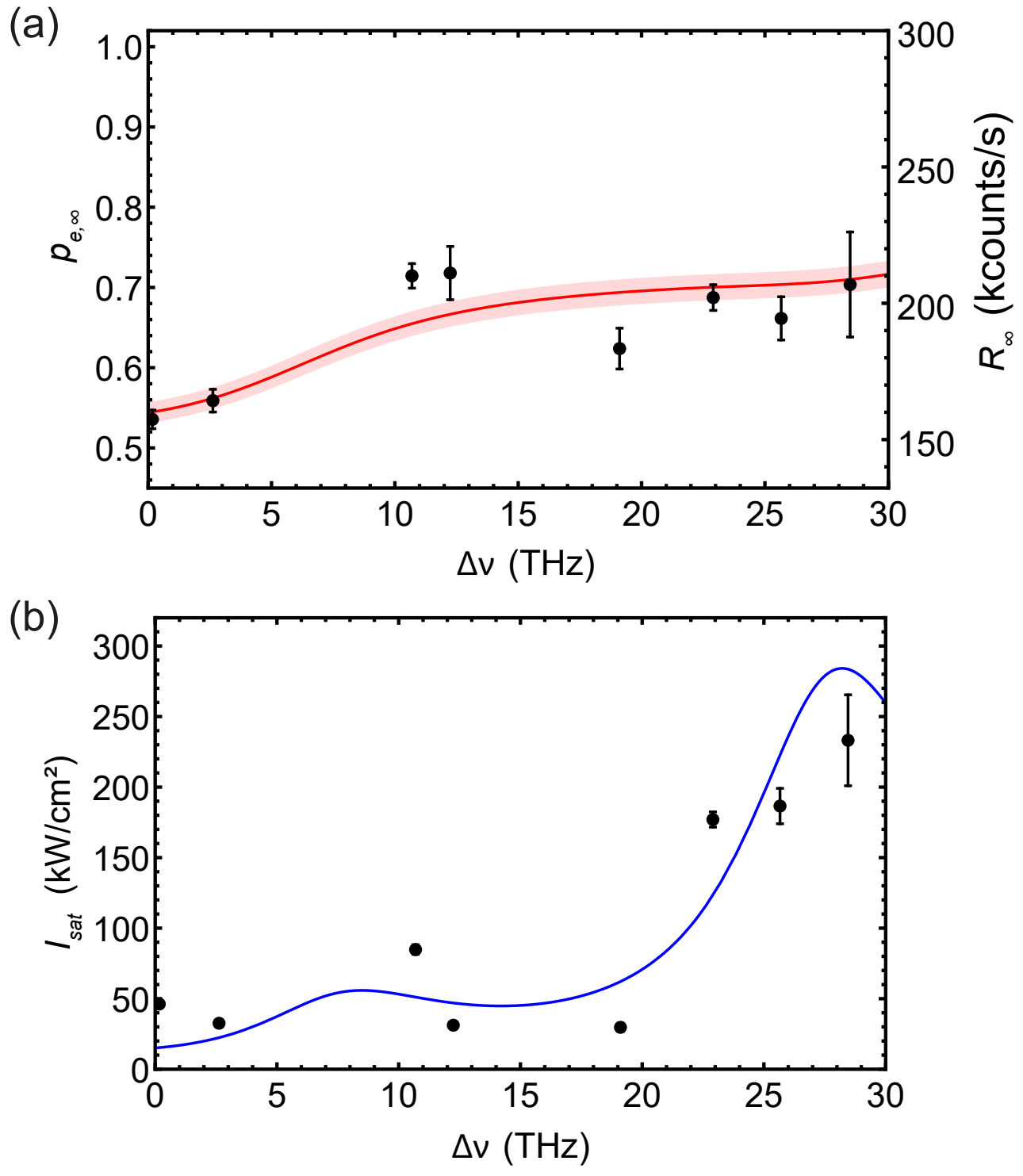

Figure 5. (a) Maximum rate $R_\infty$ of photons detected at room temperature, and corresponding maximum excited state population $p_{e,\infty}$, plotted against blue detuning of excitation laser from ZPL. Red line: room temperature $p_{e,\infty}$ as calculated using Eq.(3). Shading: region covered by one standard deviation of the fitted efficiency $\alpha$ for collection and detection. (b) Measured saturation intensity $I_{\text{sat}}$ plotted against laser frequency detuning from the ZPL. The expected saturation intensity dependence on detuning for a 7 $\mu$m thick anthracene crystal is shown as a solid line. Supplementary Information discusses the spread of data points around this curve.

The same data also provide information on the variation of $I_{\text{sat}}$ with detuning, as plotted by the data points in Fig. 5(b). Figure 2 shows that $I_{\text{sat}}$ at room temper-ature should increase monotonically with detuning, but this is not the behaviour in Fig. 5(b). The reason is that we measure (and plot) the intensity $I_{ext}$ incident on the surface of the sample, whereas $I$ in the theory refers to the intensity at the site of the DBT molecule itself. The ratio of these two is frequency dependent because the sample has several parallel dielectric interfaces making a stack of etalons. We know that the anthracene crystal on the glass substrate is more than 5 $\mu$m thick and that the PVA spin-coated top layer is 100 nm thick. A simple model with these two layers reproduces the structure seen in Fig. 5(b) when we set the anthracene thickness to be 7 $\mu$m, as shown by the blue curve. If we also include the 120 $\mu$m-thick glass substrate, the additional fine fringes provide an explanation for the scatter of the data points around the line in Fig. 5(b) (see Supplementary Information).

[1] Note, however, that the value of $\alpha$ varies by as much as 50% from one molecule to another. We think this is due to variation in the depth of the molecule within the crystal.

## IV. DBT MOLECULE AT 77 K

The room-temperature limit of $p_{e,\infty} \leq 0.75$ is not high enough to make a good triggered photon source, but a significant improvement is possible if $\Gamma 2_n$ can be reduced. Figure 6(a) shows $p_{e,\infty}$ as a function of the blue detuning of the trigger pulse, calculated from Eq.(3) for five values of $\Gamma 2_n$. The 8.2 THz curve is the same as the red curve in Fig. 5, peaking at $p_{e,\infty} = 0.75$ near a detuning of $35 - 40$ THz. That peak is improved by reducing the dephasing rate, reaching $p_{e,\infty} = 0.99$ when $\Gamma 2_n/(2\pi) = 0.7$ THz. The spectral structure in $p_{e,\infty}$ becomes increasingly evident as $\Gamma 2_n$ is reduced because the resonances become narrower, with the dips in $p_{e,\infty}$ corresponding to the regions between resonances, where the effect of the off-resonant $R_0$ is more significant.

In order to ascertain what level of cooling is required to reach $\Gamma 2_n \leq 0.7$ THz, we placed a DBT-doped anthracene crystal in a closed-cycle cryostat (Montana Cryostation) and cooled it to the base temperature of 3.5 K. We imaged a single molecule, and measured the width of its $S_{0,0} \rightarrow S_{1,0}$ scattering resonance over a range of excitation intensities (see Supplementary Information). This gave a natural width of 38(2) MHz, corresponding to $\Gamma 2_0/(2\pi) = 19(1)$ MHz and a lifetime of 4.2(2) ns. We then increased the temperature to 77 K, pumped the molecule on the strong $n = 7$ transition 8.7 THz to the blue of the ZPL, and dispersed the fluoresence on a single-photon sensitive spectrometer (Andor Shamrock with a Newton EMCCD). This gave the spectrum shown by the data points in Fig. 6(b), after subtracting a background of light scattered by the anthracene. The main features of this spectrum are well reproduced by the red line, which is a plot of Eq.(8), taking $\Gamma 2_n/(2\pi) = 0.7$ THz. On the low-frequency side of the ZPL there is a small shoulder. The residuals plotted in the lower panel show that this is a sideband centred at a detuning of 2 THz. In section II of their paper [29], Grandi *et al.* identified a possible local phonon mode of this system at an energy corresponding to 40 K ($\sim 1$ THz), so we surmise that the sideband here corresponds to a two-phonon excitation [37] of that mode.

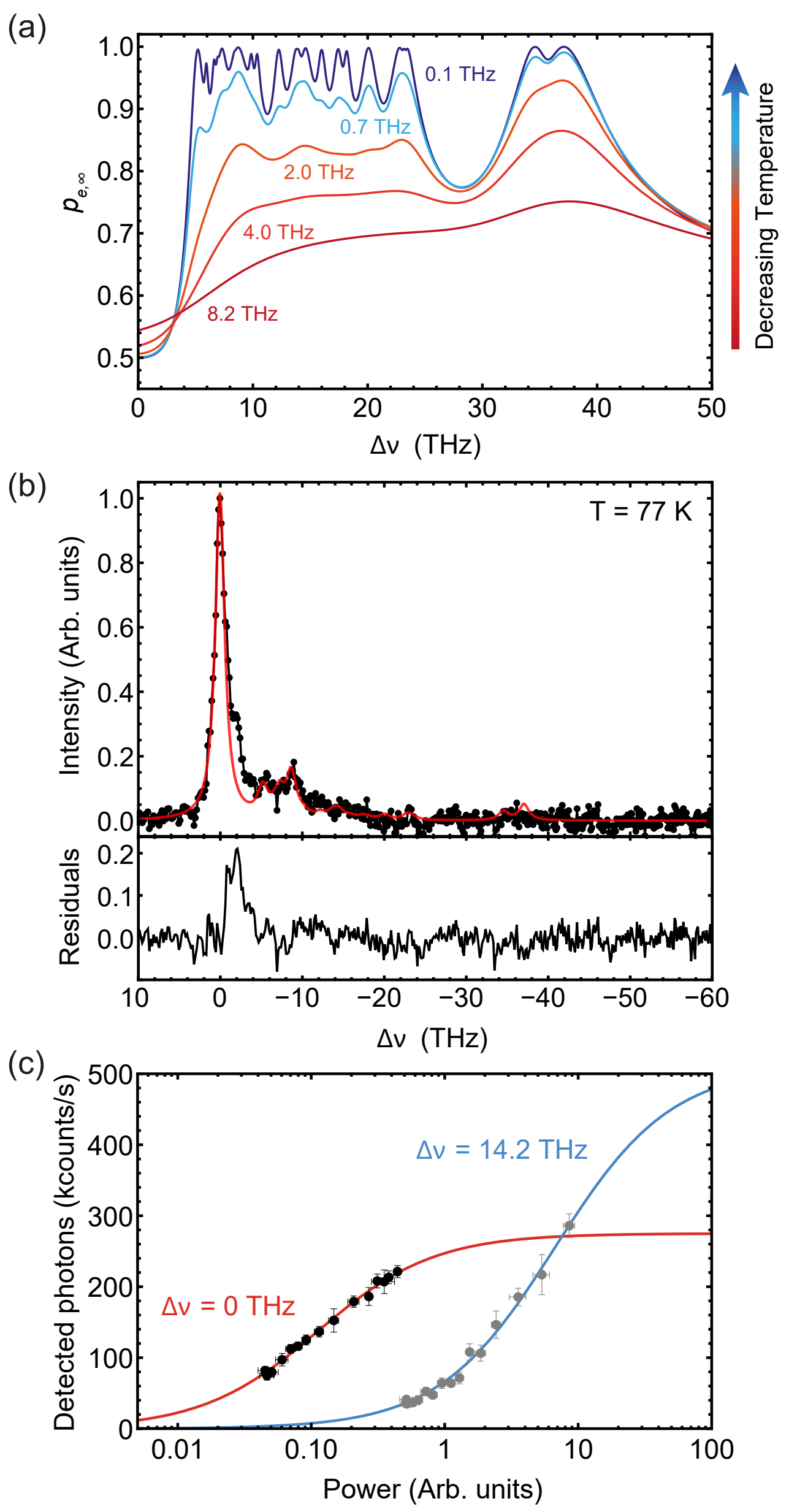

Figure 6. (a) Calculated maximum excited state probability plotted against blue detuning of the excitation laser from the ZPL, for 5 values of $\Gamma 2_n$. (b) Data points: dispersed fluorescence spectrum of a single molecule measured at a temperature of 77 K. Red line: spectrum given by Eq.(8), taking $\Gamma 2_n/(2\pi) = 0.7$ THz. Lower panel: difference between data and red line, revealing a sideband 2 THz below the ZPL. (c) Fluorescence signal *versus* excitation power with the sample cooled to 77 K, with the excitation laser on the ZPL (black dots) and tuned 14.2 THz to the blue. The lines are fits to the form $R_\infty S/(1+S)$.

Having established that liquid nitrogen temperature should be cold enough to achieve exceedingly high values of $p_e$ we put this to the test by recording the saturation of the scattering rate, as shown in Fig. 6(c). The black data points show the photon count rate in the detector as a function of the power incident on the sample, with the excitation laser tuned to the ZPL. The red line is a fit to these points of the function $R_\infty S/(1+S)$, giving $R_\infty = 275(2)$ kcounts/s. The grey data points show the photon count rates when the excitation is tuned 14.2 THz to the blue of the ZPL, where the blue line is a fit to the same function as before, giving $R_\infty = 510(4)$ kcounts/s. We expect from Eq.(3) when $\Gamma 2_n/(2\pi) = 0.7$ THz that $p_{e,\infty} = 0.50$ when pumping on the ZPL. Using this and comparing the two saturated count rates we calculate that $p_{e,\infty} = 0.93(1)$ for the blue detuned case, in agreement with the value 0.94 predicted from Eq.(3).

These results give us confidence that photons can be triggered with high efficiency using DBT in anthracene once it is cooled to liquid nitrogen temperature.

## V. CONCLUSION AND PERSPECTIVE

We have modelled the efficiency of triggering photon emission from a single PAH molecule, taking into account the thermal dephasing of the optical dipole. We find that the trigger pulse can produce a steady excited-state population $p_e$ close to 100%, but only when the stimulated emission on the ZPL is negligible. This requires the excitation laser frequency to be far detuned to the blue of the ZPL in comparison with the dephasing rate $\Gamma 2_0$.

In the case of DBT in anthracene we measure at room temperature that $\Gamma 2_n/(2\pi) = 8.2$ THz, and hence conclude that the photon generation efficiency cannot exceed 75%. We have confirmed this by isolating a single DBT molecule and investigating the saturation of the scattering rate for several frequencies of the pump laser.

By contrast, we expect that this efficiency can exceed 99% if $\Gamma 2_n/(2\pi)$ can be reduced to 0.7 THz. We have shown experimentally that this ambition can be achieved simply by cooling the sample to liquid nitrogen temperature. That is a significant step towards a fast, reliable, triggered single photon source because it is much more straightforward to operate a practical device with liquid nitrogen than with liquid helium. To our knowledge, no other types of solid state source can offer such high photon generation efficiency at 77 K. However, other PAH molecules, most notably terrylene [24], have shown high emission rates even at room temperature. As terrylene has a lower branching ratio to the ZPL than DBT, it can reach a higher $p_{e,\infty}$ of 96% at room temperature and may well be able to exceed $p_e > 99\%$ with only modest cooling (see Supplementary Information). In order to make the best use of an emitter with such high excitation probability, it is also necessary to collect the photons efficiently, and we have not addressed that challenge here. However, it has already been demonstrated that light from a dipole source can be collected very efficiently by coupling it to a resonant optical structure [13–16, 18, 24, 31, 32], and this is what we plan to do next.

## ACKNOWLEDGMENTS AND FUNDING

We are indebted to Jon Dyne, Giovanni Marinara, and Valerijus Gerulis for their expert mechanical and electrical workshop support. This work was supported in the UK by EPSRC (EP/P030130/1, EP/P01058X/1, and EP/G037043/1), dstl (DSTLX1000092512), and the Royal Society (RP110002, UF160475), and by a European Commission Marie Skłodowska Curie Individual Fellowship (Q-MoPS, 661191).

See Supplementary Information for supporting content.

[1] J. L. O'Brien, A. Furusawa, and J. Vučković, Nature Photonics **3**, 687 (2009).
[2] J. Sabines-Chesterking, R. Whittaker, S. K. Joshi, P. M. Birchall, P. A. Moreau, A. McMillan, H. V. Cable, J. L. O'Brien, J. G. Rarity, and J. C. F. Matthews, Phys. Rev. Applied **8**, 014016 (2017).
[3] N. Gisin and R. Thew, Nat. Photon. **1**, 165 (2007).
[4] J. L. O'Brien, Science **318**, 1567 (2007).
[5] J. Nunn, N. K. Langford, W. S. Kolthammer, T. F. M. Champion, M. R. Sprague, P. S. Michelberger, X.-M. Jin, D. G. England, and I. A. Walmsley, Phys. Rev. Lett. **110**, 133601 (2013).
[6] M. Collins, C. Xiong, I. Rey, T. Vo, J. He, S. Shahnia, C. Reardon, T. Krauss, M. Steel, A. Clark, and B. Eggleton, Nat. Commun. **4**, 2582 (2013).
[7] T. Meany, L. A. Ngah, M. J. Collins, A. S. Clark, R. J. Williams, B. J. Eggleton, M. J. Steel, M. J. Withford, O. Alibart, and S. Tanzilli, Laser & Photonics Reviews **8**, L42 (2014).
[8] G. J. Mendoza, R. Santagati, J. Munns, E. Hemsley, M. Piekarek, E. Martín-López, G. D. Marshall, D. Bonneau, M. G. Thompson, and J. L. O'Brien, Optica **3**, 127 (2016).
[9] R. J. A. Francis-Jones, R. A. Hoggarth, and P. J. Mosley, Optica **3**, 1270 (2016).
[10] M. Mücke, J. Bochmann, C. Hahn, A. Neuzner, C. Nölleke, A. Reiserer, G. Rempe, and S. Ritter, Phys. Rev. A **87**, 063805 (2013).
[11] D. B. Higginbottom, L. Slodička, G. Araneda, L. Lachman, R. Filip, M. Hennrich, and R. Blatt, New Journal of Physics **18**, 093038 (2016).
[12] I. Aharonovich, D. Englund, and M. Toth, Nature Photonics **10**, 631 (2016).
[13] X. Ding, Y. He, Z.-C. Duan, N. Gregersen, M.-C. Chen, S. Unsleber, S. Maier, C. Schneider, M. Kamp, S. Höfling, C.-Y. Lu, and J.-W. Pan, Physical Review Letters **116**, 020401 (2016).
[14] H. Wang, Z.-C. Duan, Y.-H. Li, S. Chen, J.-P. Li, Y.-M. He, M.-C. Chen, Y. He, X. Ding, C.-Z. Peng, C. Schneider, M. Kamp, S. Höfling, C.-Y. Lu, and J.-W. Pan,

Physical Review Letters **116**, 213601 (2016).

[15] J. C. Loredo, N. A. Zakaria, N. Somaschi, C. Antón, L. de Santis, V. Giesz, T. Grange, M. A. Broome, O. Gazzano, G. Coppola, I. Sagnes, A. Lemaître, A. Auffèves, P. Senellart, M. P. Almeida, and A. G. White, Optica **3**, 433 (2016).

[16] N. Somaschi, V. Giesz, L. De Santis, J. C. Loredo, M. P. Almeida, G. Hornecker, S. L. Portalupi, T. Grange, C. Antón, J. Demory, C. Gómez, I. Sagnes, N. D. Lanzillotti-Kimura, A. Lemaître, A. Auffèves, A. G. White, L. Lanco, and P. Senellart, Nature Photonics **10**, 340 (2016).

[17] C. Kurtsiefer, S. Mayer, P. Zarda, and H. Weinfurter, Physical review letters **85**, 290 (2000).

[18] J. Benedikter, H. Kaupp, T. Hümmer, Y. Liang, A. Bommer, C. Becher, A. Krueger, J. M. Smith, T. W. Hänsch, and D. Hunger, Physical Review Applied **7**, 024031 (2017).

[19] R. Kolesov, K. Xia, R. Reuter, M. Jamali, R. Stöhr, T. Inal, P. Siyushev, and J. Wrachtrup, Physical Review Letters **111**, 120502 (2013).

[20] B. Lounis and W. E. Moerner, Nature **407**, 491 (2000).

[21] B. Lounis and M. Orrit, Reports on Progress in Physics **68**, 1129 (2005).

[22] P.-A. Moreau, J. Sabines-Chesterking, R. Whittaker, S. K. Joshi, P. M. Birchall, A. McMillan, J. G. Rarity, and J. C. F. Matthews, Sci. Rep. **7**, 6256 (2017).

[23] B. C. Buchler, T. Kalkbrenner, C. Hettich, and V. Sandoghdar, Physical Review Letters **95**, 1 (2005).

[24] X.-L. Chu, S. Götzinger, and V. Sandoghdar, Nat. Photon. **11**, 58 (2017).

[25] J.-B. Trebbia, H. Ruf, P. Tamarat, and B. Lounis, Optics Express **17**, 23986 (2009).

[26] A. A. L. Nicolet, P. Bordat, C. Hofmann, M. A. Kol'chenko, B. Kozankiewicz, R. Brown, and M. Orrit, ChemPhysChem **8**, 1215 (2007).

[27] K. D. Major, Y.-H. Lien, C. Polisseni, S. Grandi, K. W. Kho, A. S. Clark, J. Hwang, and E. A. Hinds, Review of Scientific Instruments **86**, 083106 (2015).

[28] R. Loudon, *The Quantum Theory of Light*, 3rd ed. (Oxford Science Publications, 2000) p. 81.

[29] S. Grandi, K. D. Major, C. Polisseni, S. Boissier, A. S. Clark, and E. A. Hinds, Physical Review A **94**, 063839 (2016).

[30] A. A. L. Nicolet, P. Bordat, C. Hofmann, M. A. Kol'chenko, B. Kozankiewicz, R. Brown, and M. Orrit, ChemPhysChem **8**, 1929 (2007).

[31] S. Checcucci, P. E. Lombardi, S. Rizvi, F. Sgrignuoli, N. Gruhler, F. B. C. Dieleman, F. S. Cataliotti, W. H. P. Pernice, M. Agio, and C. Toninelli, Light: Science & Applications **6**, e16245 (2016).

[32] P. Lombardi, A. P. Ovvyan, S. Pazzagli, G. Mazzamuto, G. Kewes, O. Neitzke, N. Gruhler, O. Benson, W. H. P. Pernice, F. S. Cataliotti, and C. Toninelli, ACS Photonics **5**, 126 (2018).

[33] C. Toninelli, K. Early, J. Bremi, A. Renn, S. Götzinger, and V. Sandoghdar, Optics Express **18**, 6577 (2010).

[34] C. Polisseni, K. D. Major, S. Boissier, S. Grandi, A. S. Clark, and E. A. Hinds, Optics Express **24**, 5615 (2016).

[35] J. Lakowicz, *Principles of Fluorescence Spectroscopy* (Kluwer Academic / Plenum Publishers, 1999).

[36] S. Grandi, *Single quantum emitters: resonance fluorescence and emission enhancement*, Ph.D. thesis, Imperial College London (2017).

[37] W. H. Hesselink and D. A. Wiersma, The Journal of Chemical Physics **73**, 648 (1980).

# Supplementary Material: Efficient excitation of dye molecules for single photon generation

Ross C. Schofield,[1] Kyle D. Major,[1] Samuele Grandi,[1] Sebastien Boissier,[1] E. A. Hinds,[1] and Alex S. Clark[1,*]

[1]*Centre for Cold Matter, Blackett Laboratory, Imperial College London, Prince Consort Road, SW7 2AZ London, UK*

This supplementary material contains details about our confocal microscope image capture and analysis, the frequencies and strengths of the vibrational lines used to model a dibenzoterrylene (DBT) molecule in an anthracene crystal, spectroscopy of DBT cooled to cryogenic temperature, modelling of the intensity distribution within a typical sample in our microscope, and some additional modelling of the excited state population of another organic molecule/matrix, terrylene in $p$-terphenyl.

## I. CONFOCAL MICROSCOPE SCAN DATA ANALYSIS

A $10 \times 10$ micron square area was scanned over a few minutes using the confocal microscope, with a resolution of $75 \times 75$ pixels and an exposure time of 5 ms per pixel. A bright region was selected from the scan area, and $g^{(2)}(\tau)$ was measured to confirm it was an isolated dibenzoterrylene (DBT) molecule. Scans were taken for a range of pump laser frequencies and intensities, listed in Table I. Note that the intensities are derived from power meter readings taken in air after the objective lens – they are not the intensities inside the anthracene film. To each image, we fitted an elliptical Gaussian function over a 15 $\times$ 15 pixel area around the chosen molecule, to find its position, which moved over time by one or two pixels because of drifts in the optical mounts. Using these centres, we then made a global fit of Eq.(10) (main article) to all the images taken with a given laser frequency, in order to determine the $R_\infty$ and $I_{\rm sat}$ given in Table I. The errors in both $R_\infty$ and $I_{\rm sat}$ are found using a bootstrap technique, where we analyse 100 data sets which contain randomly sampled data points from the original data. From the results of this we calculate the standard deviation in the fitted parameters. Figure S1 shows the images recorded with the pump laser tuned $\Delta\nu = 19.11$ THz to the blue of the zero-phonon line, together with those given by the global fit. These values of $R_\infty$ and $I_{\rm sat}$ as a function of $\nu$ are the eight data points plotted in Fig. 5 of the main article.

## II. VIBRATIONAL LEVEL FREQUENCY DETUNING AND BRANCHING RATIOS IN DBT

Table II lists the relative frequencies and intensities of the Stokes sidebands due to the transitions $S_{1,0} \rightarrow S_{0,n}$, which we have determined from Fig. 1 in J.-B. Trebbia *et al.* [1] (Ref. 23 in the main article). These provide the values of $\omega_n$ and $\eta_n$ used in the main article.

* alex.clark@imperial.ac.uk

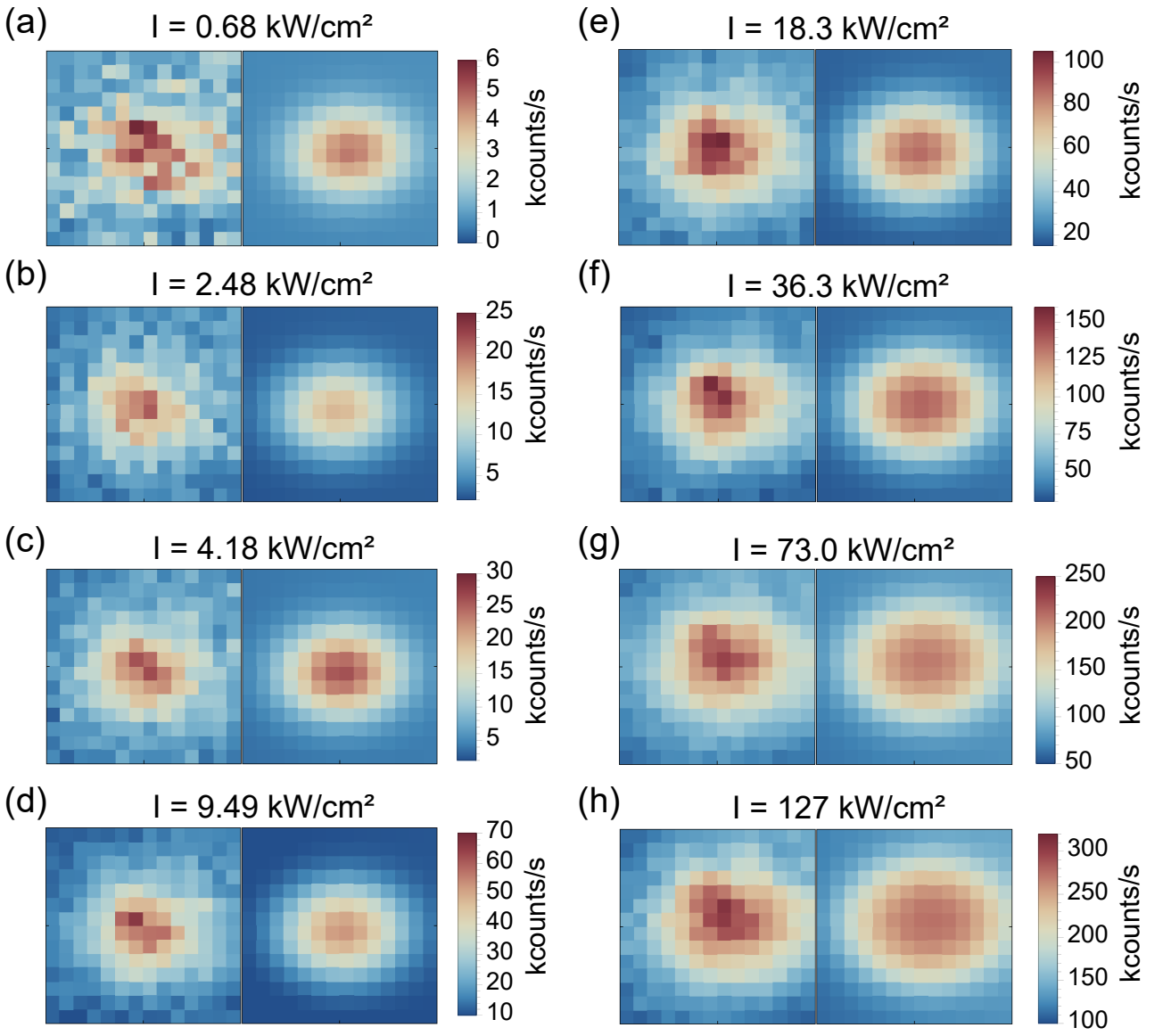

Figure S1. (a-h) Comparison of confocal microscope images (left) and fitted saturated Gaussian distributions (right) for a pump laser detuning of $\Delta\nu = 19.11$ THz. The intensities listed are obtained from the measured power incident on the sample and the known Gaussian spot size of 720 nm full width half-maximum. Each confocal scan is a 2 $\mu$m square.

## III. HOMOGENEOUS WIDTHS OF ZPL AND SIDEBANDS FOR DBT IN ANTHRACENE

We cooled a DBT-doped anthracene crystal down to 4 K in a closed-cycle cryostat (Montana Cryostation) and were able to locate single DBT molecules using confocal microscopy. We then scanned the frequency of a Ti:Sapphire laser across the $S_{0,0} \rightarrow S_{1,0}$ ZPL transition for three illumination powers. While scanning, we monitored the red-shifted fluorescence on a single-photon detector. The results, shown in Fig. S2(a), have the expected Lorentzian lineshape. The width of this scattering resonance, plotted in Fig. S2(b), should broaden with increasing power according to $2\pi\delta\nu = 2\,\Gamma 2_0\sqrt{1+S}$. We see that the points do indeed follow that form, extrapolating at zero intensity to a natural width of 38(2) MHz, meaning $\Gamma 2_0/(2\pi) = 19(1)$ MHz for the ZPL.

| $\Delta\nu$ (THz) | Intensity (kW/cm$^2$) | | | | | | | | | $R_\infty$ (kcounts/s) | $I_{\rm sat}$ (kW/cm$^2$) |
|---|---|---|---|---|---|---|---|---|---|---|---|
| 0.17 | 2.38 | 7.46 | 16.7 | 32.3 | 64.0 | 131 | 238 | | | 157.6(3.4) | 46(2) |
| 2.62 | 2.69 | 10.0 | 19.4 | 37.9 | 78.4 | 149 | 239 | | | 164.5(4.2) | 32(2) |
| 10.69 | 2.40 | 8.08 | 16.5 | 33.2 | 64.0 | 128 | 240 | | | 210.3(4.5) | 85(3) |
| 12.24 | 0.49 | 1.96 | 3.91 | 7.83 | 15.6 | 31.6 | 74.3 | | | 211.3(9.8) | 31(3) |
| 19.11 | 0.68 | 2.48 | 4.18 | 9.49 | 18.3 | 36.3 | 73.0 | 127 | | 183.6(7.5) | 30(3) |
| 22.90 | 1.53 | 3.36 | 6.41 | 12.2 | 24.7 | 49.2 | 99.3 | 189 | 295 | 202.3(4.7) | 177(5) |
| 25.66 | 2.96 | 11.4 | 22.7 | 46.1 | 91.2 | 182 | 281 | | | 194.7(7.9) | 187(13) |
| 28.45 | 1.26 | 4.31 | 8.41 | 16.6 | 37.8 | 75.7 | 147 | 201 | | 207.1(19.3) | 233(32) |

Table I. Determination of $R_\infty$ and $I_{\rm sat}$ for several frequencies of excitation. At each frequency we list the peak laser intensities incident on the sample and the values of $R_\infty$ and $I_{\rm sat}$ derived from the saturation of the images. Frequencies are given as detunings from the zero-phonon line at 382.16 THz.

| $\omega_n/(2\pi)$ (THz) | Relative Weight $\eta_n$ |
|---|---|
| 382.16 | $6.596\times10^{-1}$ |
| 387.36 | $5.410\times10^{-2}$ |
| 388.10 | $5.419\times10^{-3}$ |
| 388.78 | $5.414\times10^{-3}$ |
| 389.10 | $4.836\times10^{-3}$ |
| 389.46 | $4.199\times10^{-2}$ |
| 390.56 | $5.402\times10^{-3}$ |
| 390.87 | $8.346\times10^{-2}$ |
| 392.01 | $5.104\times10^{-3}$ |
| 392.52 | $4.812\times10^{-3}$ |
| 394.41 | $9.696\times10^{-3}$ |
| 395.94 | $7.669\times10^{-3}$ |
| 396.09 | $7.380\times10^{-3}$ |
| 396.74 | $1.429\times10^{-2}$ |
| 398.15 | $5.349\times10^{-3}$ |
| 399.59 | $6.203\times10^{-3}$ |
| 400.33 | $3.029\times10^{-3}$ |
| 402.24 | $9.928\times10^{-3}$ |
| 404.95 | $9.044\times10^{-3}$ |
| 405.34 | $3.280\times10^{-3}$ |
| 405.62 | $3.854\times10^{-3}$ |
| 416.67 | $6.074\times10^{-3}$ |
| 416.81 | $1.212\times10^{-2}$ |
| 419.26 | $3.198\times10^{-2}$ |

Table II. Frequencies of the lines in the fluorescence spectrum of DBT in anthracene and the relative peak heights, determined from Fig. 1 in J.-B. Trebbia *et al.* [1]. We take the zero-phonon line frequency as 382.16 THz.

We then scanned the laser frequency across the $S_{0,0} \to S_{1,7}$ transition at a wavelength of $\sim$ 767 nm, using a laser intensity well below saturation (estimated $S \simeq 0.01$), again monitoring the red-shifted fluorescence. A Lorentzian fit to the data, shown in Fig. S2(c), gives $\Gamma 2_7/(2\pi) = 23(3)$ GHz, which is not very different from the "$\sim$ 40 GHz at 2 K" given in Ref. [1]. We expect other vibrational transitions to have similar picosecond timescale decays and therefore similar widths.

## IV. INHOMOGENEOUS BROADENING OF DBT IN ANTHRACENE

Local variations within the crystal host cause the frequency of the zero-phonon line to vary from molecule to molecule. To investigate this, we cooled a DBT-doped anthracene crystal to 10 K and took confocal scans for various frequencies of excitation. At each frequency we summed the fluorescence signal from all the pixels (coming from many molecules), and this gave the inhomogeneously-broadened spectrum plotted by the data points in Fig. S2(d). We compare the line shape with a Gaussian distribution (red dashed), a Voigt distribution (blue dotted), and a Lorentzian distribution (black line), all of which have a full width at half maximum of 360 GHz. The Lorentzian represents the data best, fitting well on the red wing of the line. On the blue side, the data points have a broad wing that may be due to molecules near the surface of the anthracene experiencing a strain-induced shift to higher frequencies. Because the 8.2 THz homogeneous width at room temperature is so much greater than this inhomogeneous distribution, we could not determine the local frequency shift of the molecule used for Fig. 5 of the main article. We therefore took it to be 382.16 THz – the frequency of the highest point in Fig. S2(d).

## V. INTENSITY INSIDE THE SAMPLE

The sample in our experiment comprised a glass cover slip 120 $\mu$m thick, a doped anthracene layer, and a 100 nm thick top layer of polyvinyl alcohol (PVA). Our atomic force microscope could not measure the thickness of the anthracene as it was more than 5 $\mu$m. In Fig. 5(b) of the main article we show that $I_{\rm sat}$ oscillates with pump laser frequency, which we understand by modelling the etalons formed in the anthracene and PVA. The $\sim$ 10 THz period of the oscillations indicates that the anthracene crystal is $\sim$ 7 $\mu$m thick. Our model propagates the light with Fresnel reflection and transmission at each interface in the sample stack. The input light passes from a half-space of air (refractive index $n = 1$), to the PVA layer ($n = 1.45$), then the anthracene crystal ($n = 1.8$ along the crystal axis where we expect molecules to be aligned), and finally exits into a glass substrate ($n = 1.45$) which forms the other half-space. The predicted $I_{sat}$ from our model – a monotonically increasing function of detuning – is then scaled by intensity ratio inside and outside the stack to yield the modulated curve in Fig. 5(b) of the main article. When we add the glass substrate as a third etalon in the model with a final half-space of air, we see an additional fast modulation, which is plotted in Fig. S3. We do not know the optical thickness of the glass accurately enough to trust the phase of the fast oscillations, but the depth of this modulation is reasonably consistent with the excess spread of the data points around the curve in Fig. 5(b) of the main article. We think this provides a probable explanation for that spread.

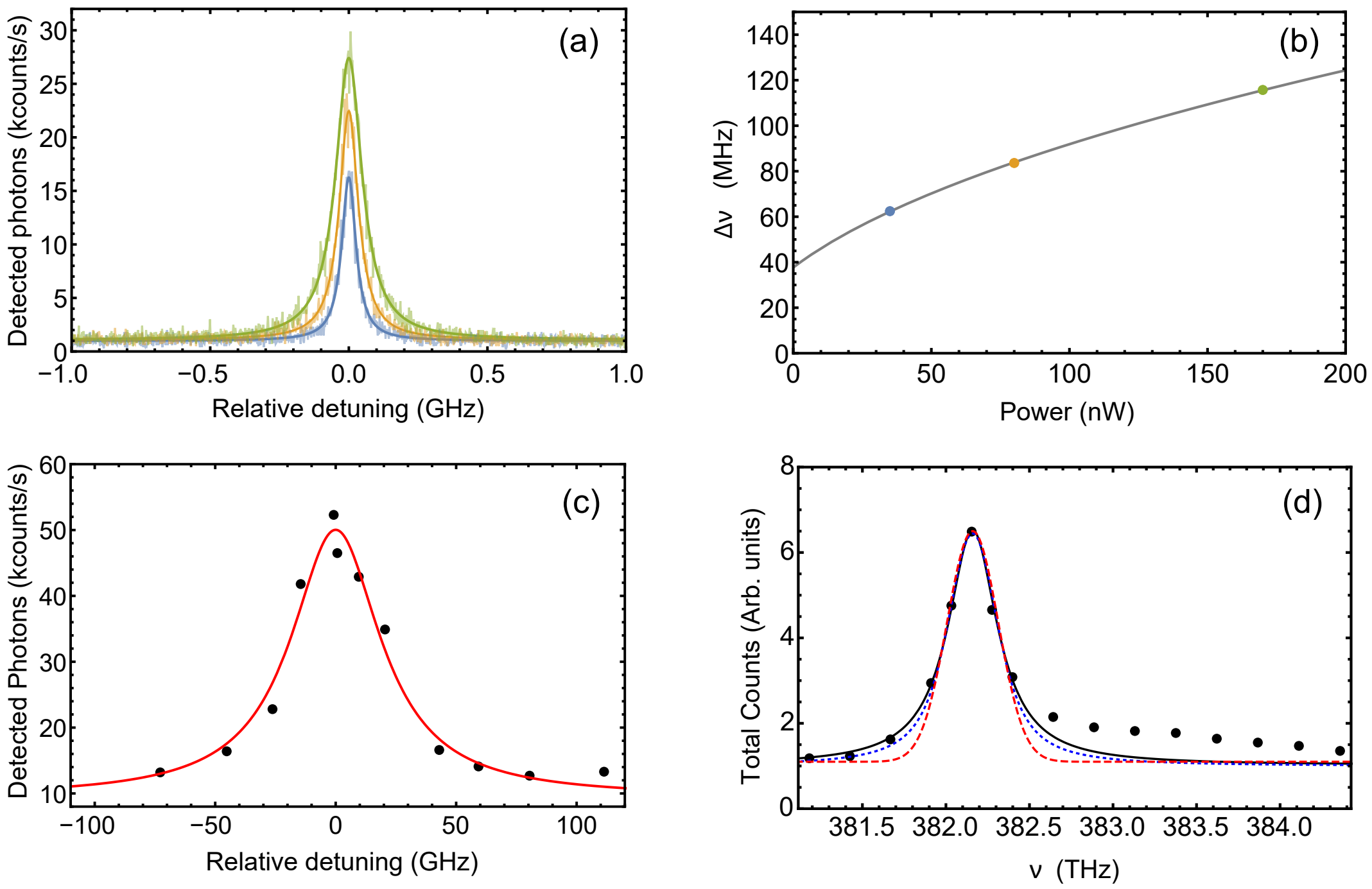

Figure S2. (a) Red-shifted fluorescence induced by scanning the excitation laser across the zero phonon line of a DBT molecule at 4 K temperature. We use three different pump laser powers: 35 nW (blue), 80 nW (orange) and 170 nW (green). (b) Power broadening of the ZPL in DBT, with widths from fits in (a). (c) Scattered red shifted photons detected while scanning a laser frequency across the $S_{0,0} \rightarrow S_{1,7}$ vibrational level transition. The fit is a Lorentzian with a full-with at half maximum of 45(6) GHz. (d) Total fluorescence from confocal scans plotted for various laser frequencies. Fits are described in the text.

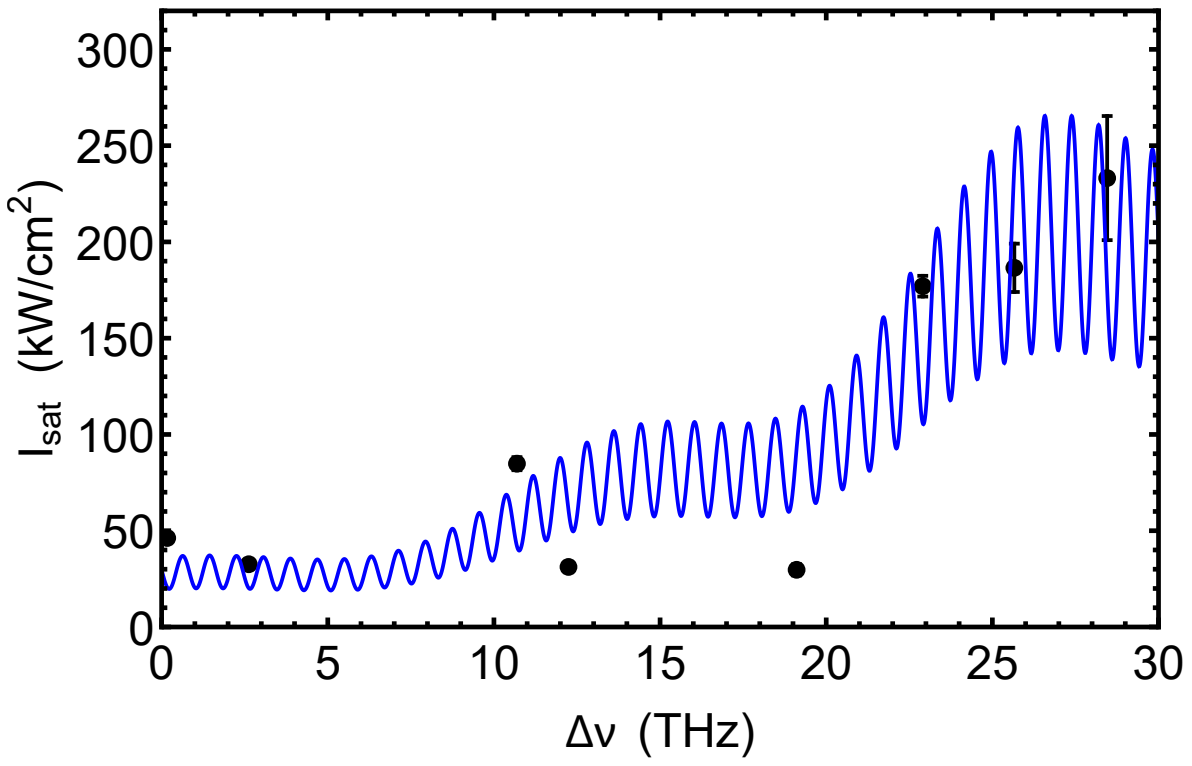

Figure S3. Change in $I_{sat}$ with frequency detuning from experiment (black points) and from our model (blue line) including modulation of intensity inside the anthracene crystal from various boundary reflections and interference between forward and backward going fields.

## VI. MODELLING THE POPULATION OF TERRYLENE AT ROOM TEMPERATURE

We have also applied our model to Terrylene (Tr) in *p*-terphenyl – another organic molecule emitter-host combination that has been investigated at both low temperature [2–4] and room temperature [5–7]. Recently Chu *et al.* [7] demonstrated an overall source efficiency, including detection, of 68% for Tr in *p*-terphenyl, pumped with a detuning of $\sim$ 42 THz. We have estimated the strengths and frequencies of the fluorescence lines of Tr out to 55 THz detuning, using the cold spectra in Fig. 1 and Fig. 2 of Ref. [8]. On broadening these spectra to match Fig. 3 of Ref. [6] we find a room temperature dephasing rate of $\Gamma 2 \simeq 6$ THz, similar to the 8.2 THz of DBT. Applying the analysis of our main article to this new system at room temperature, we calculate the maximum excited state population $p_{e,\infty}$ plotted in Fig. S4. When the trigger pulse is 40 THz detuned from the ZPL, the steady-state excited population is as high as 96% – significantly more than DBT can provide at room temperature. This should increase to more than 99% if $\Gamma 2$ can be reduced to 2 THz. The large $p_e$ is due to the weakness of the ZPL relative to its vibrational sidebands, which is beneficial for reaching a high degree of excitation, but unfavourable for yielding an output of identical photons having the ZPL frequency. Tr is also disadvantaged with respect to DBT as it has a higher inter-system crossing probability to the triplet state of $\sim 10^{-4}$ and a longer triplet lifetime of $\sim 170\,\mu$s [7], reducing the yield of photons emitted on the desired singlet transition or requiring a low repetition rate of pulsed excitation.

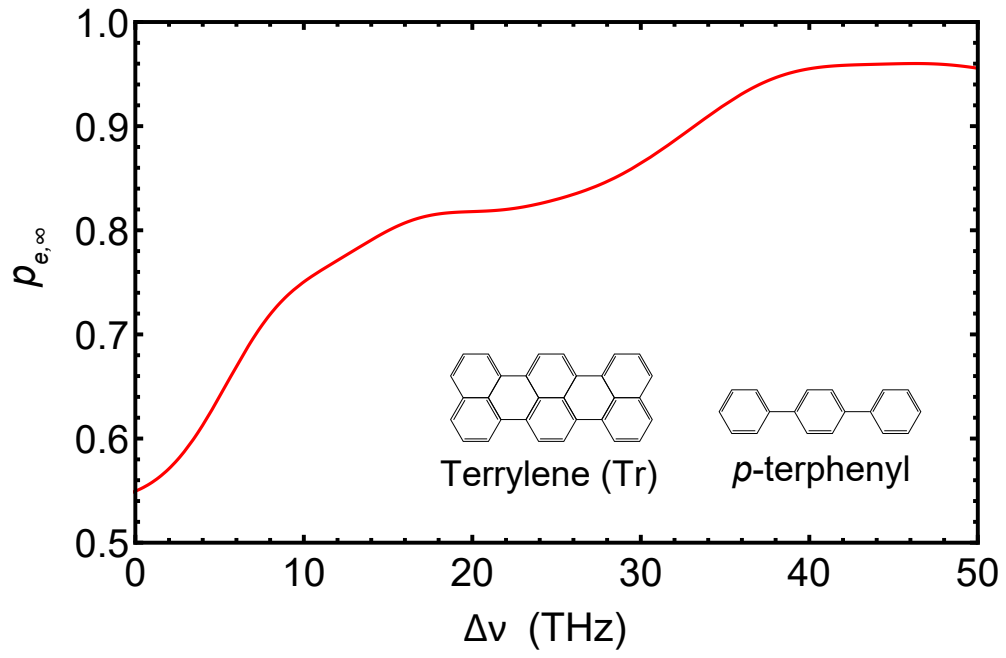

Figure S4. The maximum achievable excited state population $p_{e,\infty}$ of terrylene in $p$-terphenyl at room temperature. Their molecular structure is inset.

[1] J.-B. Trebbia, H. Ruf, P. Tamarat, and B. Lounis, Optics Express **17**, 23986 (2009).
[2] S. Kummer, T. Basché, and C. Bräuchle, Chemical Physics Letters **232**, 414 (1995).
[3] P. Tamarat, A. Maali, B. Lounis, and M. Orrit, The Journal of Physical Chemistry A **104**, 1 (2000).
[4] B. C. Buchler, T. Kalkbrenner, C. Hettich, and V. Sandoghdar, Physical Review Letters **95**, 1 (2005).
[5] L. Fleury, B. Sick, G. Zumofen, B. Hecht, and U. P. Wild, Molecular Physics **95**, 1333 (1998).
[6] F. Kulzer, F. Koberling, T. Christ, A. Mews, and T. Basché, Chemical Physics **247**, 23 (1999).
[7] X.-L. Chu, S. Götzinger, and V. Sandoghdar, Nat. Photon. **11**, 58 (2017).
[8] S. Kummer, F. Kulzer, R. Kettner, T. Basché, C. Tietz, C. Glowatz, and C. Kryschi, The Journal of Chemical Physics **107**, 7673 (1997).